\documentclass[prc,showpacs,showkeys,superscriptaddress,nofootinbib,twocolumn,floatfix]{revtex4}
\usepackage{amsfonts}
\usepackage{graphicx,color,amsmath,amssymb}

\def\blfootnote{\xdef\@thefnmark{}\@footnotetext}

\begin{document}

\title{Hadro-chemistry effects on charm decayed leptons in heavy-ion collisions}
\author{Kai Lu}
\affiliation{Department of Applied Physics, Nanjing University of Science and Technology, Nanjing 210094, China}
\author{Min He}
\affiliation{Department of Applied Physics, Nanjing University of Science and Technology, Nanjing 210094, China}

\date{\today}

\begin{abstract}
Charm-hadrons possess versatile hadro-chemistry as characterized by various transverse-momentum-dependent ratios between their different species.
In particular, the charm hadro-chemistry may be modified in relativistic heavy-ion collisions with respect to proton-proton collisions at the same
energy, as caused by novel diffusion and hadronization mechanisms of charm quarks in the environment of the created quark-gluon plasma (QGP) in the former.
Inspired by recent measurements of leptons from charm-hadron decays (separated from bottom decays) in Pb-Pb and Au-Au collisions, we investigate
the effects of the charm hadro-chemistry on the leptonic observables. We find that full consideration of charm hadro-chemistry in both proton-proton
and heavy-ion collisions causes only mild change of charm-leptons' suppression factor with respect to previous calculations hadronizing charm quarks
into $D$ mesons only, whereas the resulting change (increase) in the charm-leptons' elliptic
flow turns out to be more pronounced as a consequence of the larger collectivity of $\Lambda_c$ baryons than $D$ mesons.
\end{abstract}

\pacs{25.75.Dw, 12.38.Mh, 25.75.Nq}
\keywords{Heavy Flavor, Quark Gluon Plasma, Ultrarelativistic Heavy-Ion Collisions}

\maketitle

\section{Introduction}
\label{sec_intro}
Heavy-ion collisions at collider energies produce a novel state of deconfined strong-interaction matter
that is known as quark-gluon plasma (QGP) and behaves as a strongly-coupled near-perfect fluid~\cite{Shuryak:2014zxa,Busza:2018rrf}.
Heavy quarks, charm and bottom, have masses much larger than the intrinsic non-perturbative scale $\Lambda_{\rm QCD}$
as well as the typical temperatures reached in current heavy-ion collision experiments. Thus they are produced in the
initial stage of the collisions and participate in the full evolution of the system (yet with the charm/bottom number
conserved), constituting powerful probes of the matter created (for pertinent recent reviews, see~\cite{Dong:2019unq,Dong:2019byy,Zhao:2020jqu,He:2022ywp}).

Historically, the traditional observables associated with heavy flavor probes were non-photonic leptons (electrons and muons) from
semileptonic decays of combined charm and bottom hadrons~\cite{STAR:2006btx,PHENIX:2006iih,Sakai:2013ata,ALICE:2015xyt,ALICE:2016ywm,ALICE:2020sjb}. The strong
suppression and large elliptic flow of heavy flavor leptons provided evidence of strong coupling of heavy quarks with the medium~\cite{Rapp:2009my}.
In recent years, measurements of charm-hadrons have become accessible, which not only corroborated the strong collectivity of charm probes in a more direct fashion~\cite{ALICE:2015vxz,ALICE:2013olq,ALICE:2017pbx,ALICE:2021rxa,CMS:2017qjw,CMS:2017vhp,STAR:2018zdy,STAR:2017kkh},but also revealed versatile charm hadro-chemistry
as characterized by various transverse-momentum ($p_T$) dependent charm-hadron ratios~\cite{ ALICE:2015dry,ALICE:2021kfc,ALICE:2021bib,CMS:2019uws,STAR:2021tte,STAR:2019ank}.
Interpretation of the strong collectivity and differential hadro-chemistry of charm-hadrons requires inputs of charm quark diffusion in the quark-gluon plasma with a rather small spatial diffusion coefficient and hadronization in terms of recombination that plays the dominant role in the low and intermediate $p_T$ regime~\cite{He:2012df,He:2019vgs,Plumari:2017ntm,Nahrgang:2013xaa,Ke:2018tsh,Katz:2019fkc,Cao:2019iqs,Beraudo:2022dpz}.

Going beyond the early measurements of comprehensive heavy flavor leptons, leptons from charm decays have been rather recently separated from those from bottom decays, both in Pb-Pb collisions at $\sqrt{s_{\rm NN}}=5.02$\,TeV by ATLAS collaboration~\cite{ATLAS:2020yxw,ATLAS:2021xtw} and in Au-Au collisions at $\sqrt{s_{\rm NN}}=200$\,GeV by PHENIX~\cite{PHENIX:2015ynp} and STAR collaboration~\cite{STAR:2021uzu,Si:2019esg}, offering an opportunity to test if differential charm hadro-chemistry leaves any imprints on the charm leptonic observables. Indeed in previous studies of heavy flavor leptons~\cite{vanHees:2005wb,Gossiaux:2008jv,He:2011qa,He:2014cla,Song:2016rzw,Katz:2019fkc}, only $D$ mesons were accounted for in the charm quark hadronization and in the ensuing charm decayed leptons. However, from the latest charm hadro-chemistry point of view~\cite{He:2019vgs,Andronic:2021erx}, charm quarks are hadronized into different species of charm-hadrons that possess varying $p_T$ spectra (including their anisotropy) and have different kinematics and branching ratios when decaying into leptons, leading to possible changes in the charm leptonic observables with respect to the case of hadronizing all charm quarks into $D$ mesons only.

The aim of the present work is to investigate the quantitative effects of full charm hadro-chemistry on the suppression and collective flow of charm leptons. A first attempt in this regard was made in Refs.~\cite{Sorensen:2005sm,Martinez-Garcia:2007mzr}, where a schematic charm-baryon-to-meson enhancement in Au-Au collisions (relative to $pp$ collisions) was assumed without taking account of realistic energy loss, and an additional moderate suppression of charm electrons at intermediate $p_T$ was estimated as a result of the smaller branching ratio of $\Lambda_c$ baryons decaying to electrons than $D$ mesons. In the present work, we employ the full $p_T$ dependent charm hadro-chemistry computed in a comprehensive charm transport approach recently developed in~\cite{He:2019vgs} for heavy-ion collisions. For calculating the charm leptons spectrum in $pp$ collisions as the baseline to characterize the medium effect, we use the $p_T$ spectra of various charm-hadrons computed in an extended statistical hadronization model~\cite{He:2019tik}, which successfully explained the charm hadro-chemistry measured in high-energy $pp$ collisions by ALICE collaboration~\cite{ALICE:2020wfu,ALICE:2021dhb}. We find that, with respect to calculations considering charm quarks hadronizing into only $D$ mesons, full consideration of charm hadro-chemistry in both $pp$ and heavy-ion collisions results in only mild change of charm-leptons' nuclear modification factor, while the resulting change (increase) in the charm-leptons' elliptic flow turns out to be more pronounced, which seems to be supported by the recent measurement of charm muons by the ATLAS collaboration~\cite{ATLAS:2020yxw,ATLAS:2021xtw}.

The organization of our article is as follows. In Sec.~\ref{sec_hadro_chemistry}, we briefly recall the essential ingredients and features of our models developed recently for calculating the charm hadro-chemistry in $pp$ and heavy-ion collisions and show the pertinent results in $\sqrt{s_{\rm NN}}=5.02$\,TeV semicentral Pb-Pb collisions from an updated calculation. In Sec.~\ref{sec_hadron_vs_leptons}, we perform the semileptonic decays of various charm-hadrons and compare the corresponding observables between charm muons and their parent hadrons for each species. In Sec.~\ref{sec_total_charm_leptons_vs_data}, we compare the nuclear modification factor and elliptic flow of total charm leptons computed from all charm-hadron species with the ones calculated from considering charm quarks hadronizing into only $D$ mesons. The quantitative influence of charm hadro-chemistry on charm leptonic observables is then discussed when confronted with pertinent measurements in Pb-Pb and Au-Au collisions. We finally summarize in Sec.\ref{sec_summary}.

\section{Charm hadro-chemistry in $pp$ and heavy-ion collisions}
\label{sec_hadro_chemistry}
Recent measurements of charm-hadrons differential yields in $\sqrt{s}=5.02$\,TeV $pp$ collisions at mid-rapidity by the ALICE collaboration demonstrated that charm fragmentation is non-universal across different colliding systems, as highlighted by the significant enhancement of the $\Lambda_c^+/D^0$ ratio at low $p_T$ relative to $e^+e^-$ collisions~\cite{ALICE:2020wfu}. This was successfully explained by an extended statistical hadronization model calculation~\cite{He:2019tik}, where the PDG list of charm-baryons was augmented with many more not-yet-observed states predicted from relativist quark model calculations~\cite{Ebert:2011kk}. The $p_T$ spectra of ground state $D^0$, $D^+$, $D_s^+$ and $\Lambda_c^+$ were well reproduced and the total charm cross section per unity rapidity (at mid-rapidity) was fitted to be $d\sigma^{c{\bar c}}/dy\sim 1$\,mb, which is consistent with the measured value reported in~\cite{ALICE:2021dhb}. These $p_T$ differential cross sections for the ground state charm-hadrons will be used to perform semileptonic decays, in order to get the baseline spectra for calculating the nuclear modification factors of the charm leptons in $\sqrt{s}=5.02$\,TeV Pb-Pb collisions.

\begin{figure} [!t]
\includegraphics[width=0.99\columnwidth]{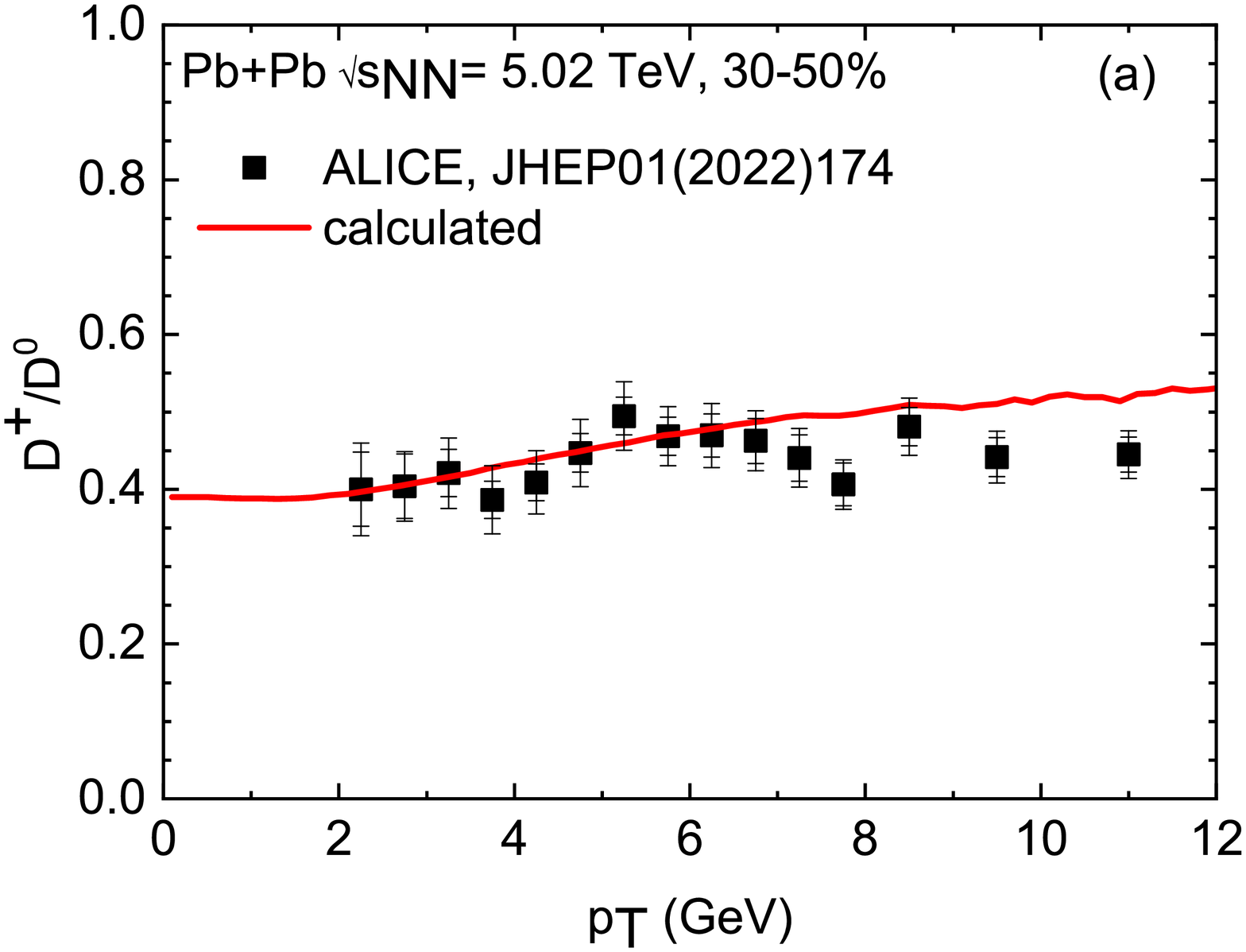}
\vspace{-0.15cm}
\includegraphics[width=0.99\columnwidth]{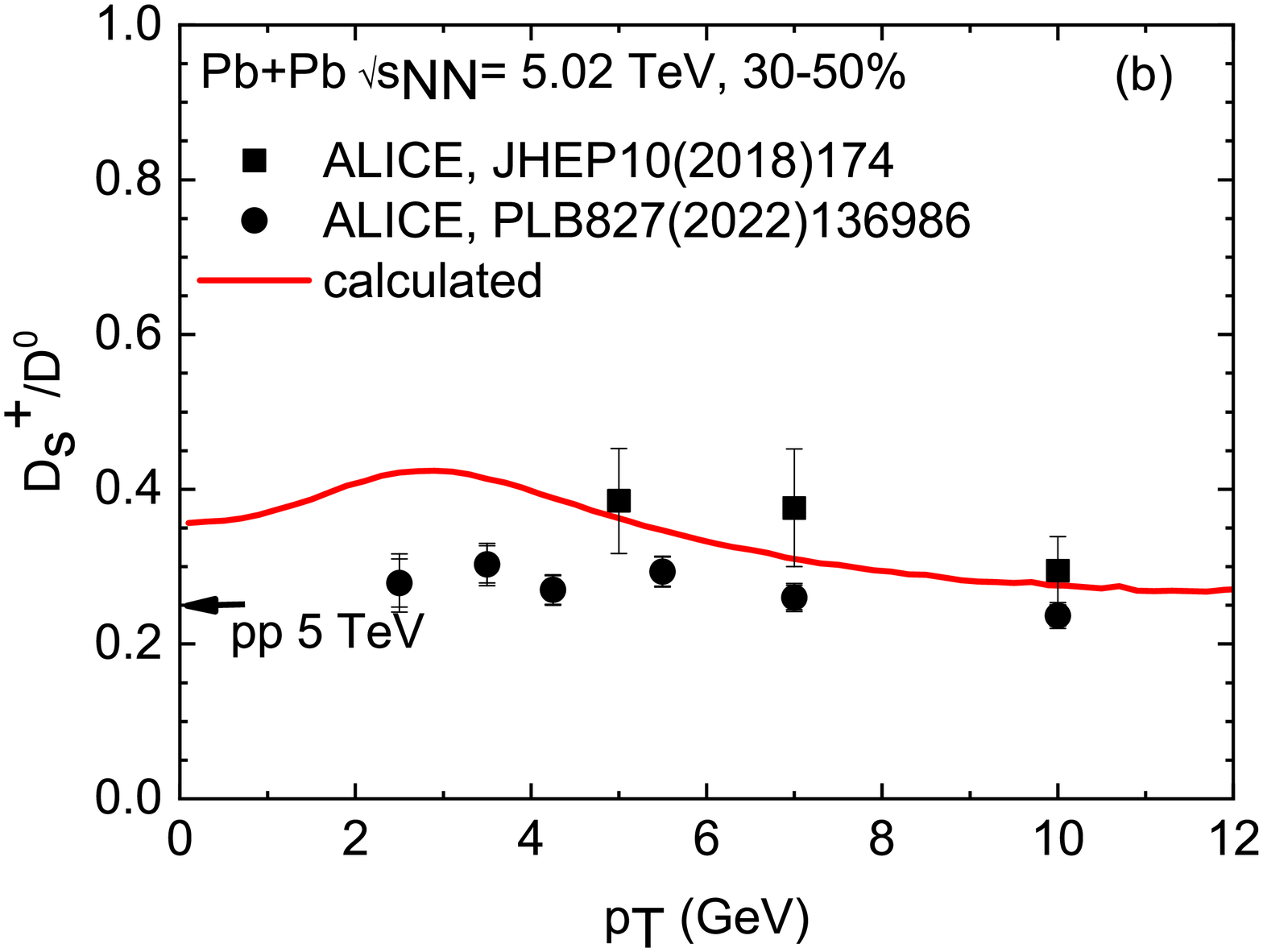}
\vspace{-0.15cm}
\includegraphics[width=0.99\columnwidth]{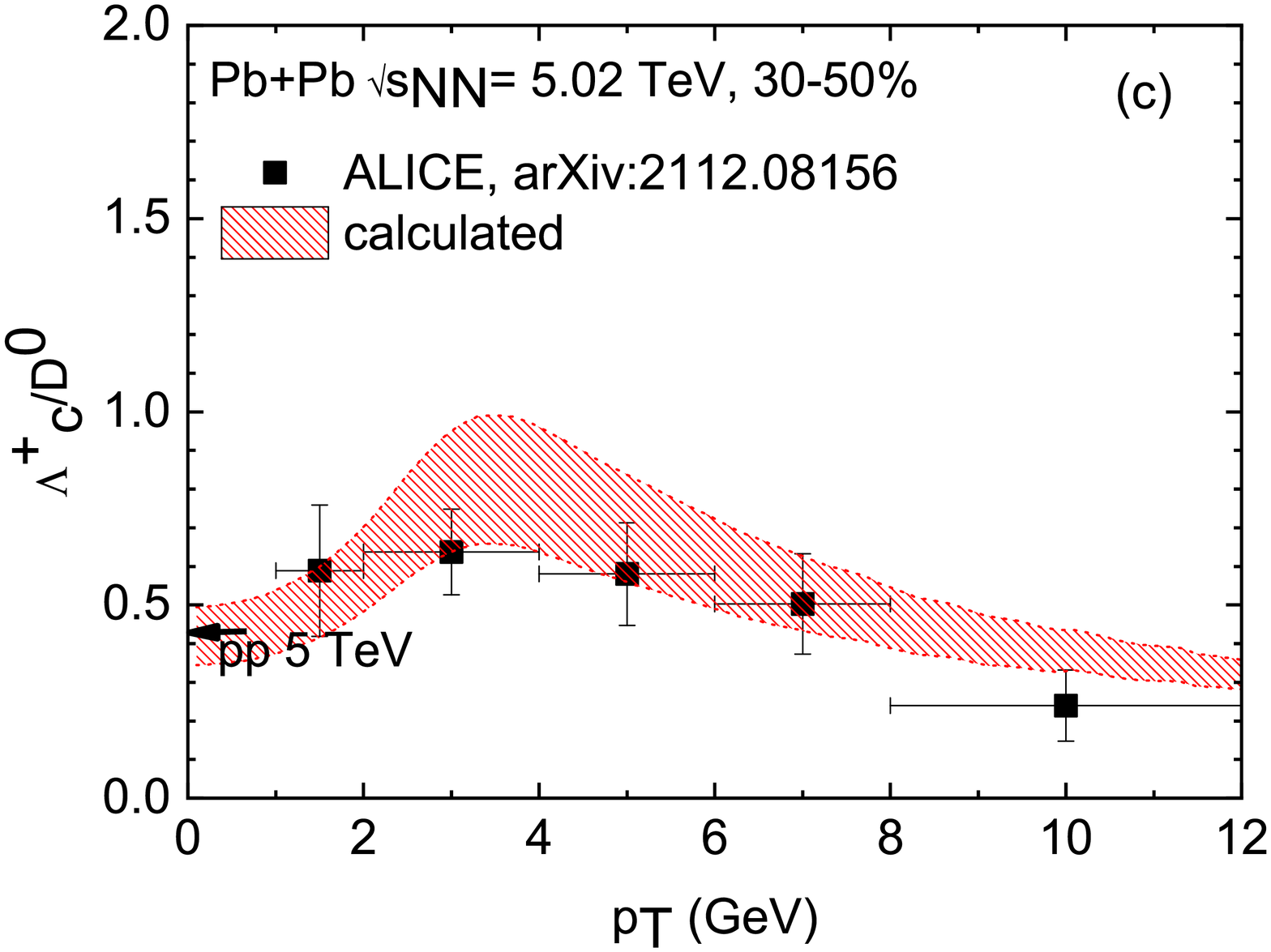}
\vspace{-0.15cm}
\caption{(Color online) Calculated charm-hadron ratios in $\sqrt{s}=5.02$\,TeV Pb-Pb collisions in the 30-50\% centrality bin at mid-rapidity in comparison with ALICE measurements~\cite{ALICE:2021rxa,ALICE:2021kfc,ALICE:2021bib}. For $\Lambda_c^+/D^0$, the band encompasses uncertainty in the branching raitos of the not-yet-measured excited charm-baryons feeding down to the ground state $\Lambda_c^+$, taken to be 50\%-100\%, cf.~\cite{He:2019vgs}.}
\label{Charm-hadron_ratios}
\end{figure}

For the charm hadro-chemistry in heavy-ion collisions, we conduct a calculation within our recently developed transport approach~\cite{He:2019vgs}. In this approach, charm quark diffusion in the hydrodynamically expanding QGP was simulated with the transport coefficient calculated in the lattice-constrained $T$-matrix approach~\cite{Riek:2010fk,Liu:2018syc}. The charm quark hadronization was modelled by resonance recombination; in particular the three-body resonance recombination model (RRM) was developed to compute the charm-baryon formation, taking advantage of the light diquark correlations in the charm-baryon sector. In addition, a method was devised to incorporate space-momentum correlations (SMCs) between the phase space distributions of charm quarks and light quarks that are built up whether through diffusion or from hydrodynamic flow. In practice, the RRM was implemented on an event-by-event basis in combination with the relativistic Langevin simulation of charm quark diffusion in QGP. With the selfconsistely determined recomination probability as a function of charm quark momentum in the fluid rest frame, this implementation allows to conserve charm quark number in each event and satisfies both kinetic and chemical equilibrium limits when using thermal quark distributions as inputs. These features, together with the inclusion of all charm-hadron species including augmented charm-baryons (same as in $pp$), are pivotal for controlled predictions of the $p_T$ dependent charm hadro-chemistry.

In Fig.~\ref{Charm-hadron_ratios}, we show the $p_T$ dependence of ground state charm-hadron ratios $D^+/D^0$, $D_s^+/D^0$ and $\Lambda_c^+/D^0$ in comparison with ALICE data in 30-50\% Pb-Pb collisions at $\sqrt{s_{\rm NN}}=5.02$\,TeV, from a calculation within the aforementioned transport model with updated light diquark masses. More specifically, unlike the degenerate diquark masses used in~\cite{He:2019vgs}, we here use distinct scalar diquark mass ($\sim 710$\,MeV) for $\Lambda_c$ states and axial-vector diquark mass ($\sim 909$\,MeV) for $\Sigma_c$ states~\cite{Ebert:2011kk}. Such a refinement yields a better description of the $p_T$ shape of the $\Lambda_c^+/D^0$ than reported in~\cite{He:2019vgs}. Compared to the corresponding ratios in $pp$ collisions~\cite{He:2019tik}, $D^+/D^0$ remains almost the same but $D_s^+/D^0$ is significantly enhanced in the low and intermediate $p_T$ regime as a result of charm recombination in the presence of strangeness equilibration in QGP~\cite{He:2012df}. We emphasize here that it is important to correctly reproduce the $D^+/D^0$ in the present context of studying the hadro-chemistry effects on the charm leptons, since the $D^+$ semileptnic decay branching ratio is almost two and half times that of $D^0$~\cite{ParticleDataGroup:2018ovx}. For $\Lambda_c^+/D^0$, the moderate enhancement relative to the $pp$ case only shows up at intermediate $p_T$ due to recombination incorporating SMCs, whereas at low $p_T$, the value turns out to be compatible with the $pp$ value, implying the integrated ratio may not change~\cite{ALICE:2021bib}. Taken as a whole, we note that these ratios, when integrated over $p_T$, are consistent with the statistical hadronization model (SHMc) calculations~\cite{Andronic:2021erx} (albeit a smaller $\Lambda_c^+/D^0$ value found in SHMc because of less charm-baryons considered therein~\cite{Andronic:2021erx}), as dictated by the {\it relative} chemical equilibrium reached in the RRM~\cite{He:2019vgs}.

\section{Charm hadrons vs decayed leptons observables}
\label{sec_hadron_vs_leptons}
\begin{figure} [!t]
\includegraphics[width=0.99\columnwidth]{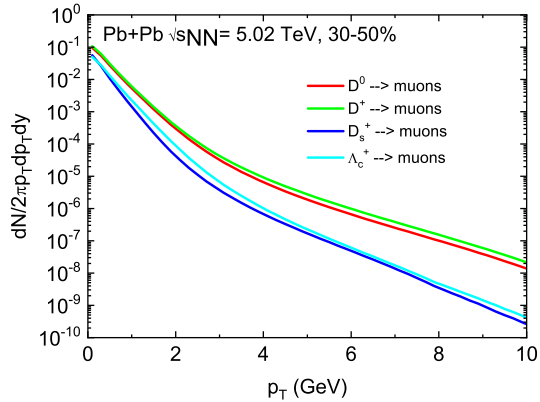}
\vspace{-0.15cm}
\caption{(Color online) Charm muons invariant $p_T$ spectra decayed from absolutely normalized $p_T$ spectra of ground state charm-hadrons $D^0$, $D^+$, $D_s^+$ and $\Lambda_c^+$ in $\sqrt{s}=5.02$\,TeV Pb-Pb collisions in the 30-50\% centrality bin at mid-rapidity. For the case of $\Lambda_c^+$, we show only the result corresponding to branching ratios $\sim 100$\% of excited $\Lambda_c$ and $\Sigma_c$ states decaying to the ground state $\Lambda_c^+$.}
\label{Charm-muons_spectra}
\end{figure}

\begin{figure} [!t]
\includegraphics[width=0.99\columnwidth]{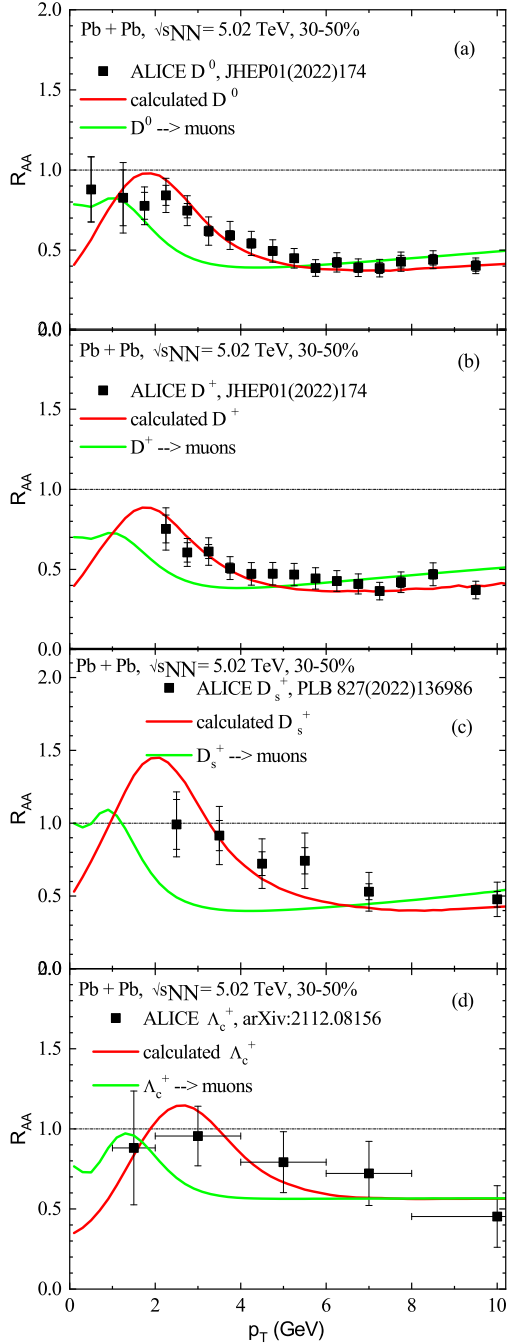}
\vspace{-0.05cm}
\caption{(Color online) $R_{\rm AA}$'s of ground state charm-hadrons in comparison their corresponding charm muons in $\sqrt{s}=5.02$\,TeV Pb-Pb collisions in the 30-50\% centrality bin at mid-rapidity. ALICE data are taken from~\cite{ALICE:2021rxa,ALICE:2021kfc,ALICE:2021bib}. For $\Lambda_c^+$, we show only the result corresponding to branching ratios $\sim 100$\% of excited $\Lambda_c$ and $\Sigma_c$ states decaying to the ground state $\Lambda_c^+$.}
\label{Charm-hadrons-muons_RAA}
\end{figure}


The calculated ground state charm-hadron $p_T$ spectra are then taken to perform semileptonic decays~\cite{He:2011qa}. More specifically, the semileptonic decays of charm-hadrons are simulated as free quark decays $c \rightarrow s + l + \bar{\nu}_l$, with the decay matrix element taken from the low-energy $V$-$A$ theory~\cite{Griffiths}: $\overline{|\mathcal{M}|^2}\propto (p_s\cdot p_{\bar{\nu}_l})(p_c\cdot p_l)$ and branching ratios taken from~\cite{ParticleDataGroup:2018ovx}. The quark masses are replaced by the corresponding hadron masses to correctly account for the phase space. The hadronic form factors have little influence on the lepton energy distribution in the parent charm-hadron rest frame~\cite{He:2011qa}. The lepton momenta are then boosted into lab frame.

In Fig.~\ref{Charm-muons_spectra}, we display the invariant muons spectra decayed from ground state $D^0$, $D^+$, $D_s^+$ and $\Lambda_c^+$ spectra that have been normalized to the realistic total charm number $dN_{c\bar{c}}/dy\sim 3.44$ (after taking account of shadowing reduction) in $\sqrt{s}=5.02$\,TeV Pb-Pb collisions in the 30-50\% centrality bin at mid-rapidity. One notes that even if the $D^+/D^0$ is around $\sim 0.4$, the invariant muons spectrum decayed from $D^+$ finally exceeds that from $D^0$, because the branching ratio of the former ($\sim 16.1\%$) is almost $2.5$ times that of the latter ($\sim 6.5\%$). The charm muons spectra decayed from $D_s^+$ and $\Lambda_c^+$ are significantly softer than those from $D^0$ and $D^+$, because of the steeper $p_T$ spectra of the former two charm-hadrons as well as kinematic effect in the decay associated with their larger masses. This, in combination with the smaller $\Lambda_c^+$ semileptonic branching ratio ($\sim 4.5\%$; the $D_s^+$ branching ratio is roughly the same as that of $D^0$), renders the total charm muons spectrum dominated by those decayed from two un-stranged charm mesons, which is more so toward high $p_T$; e.g., at $p_T\sim 1~(6)$\,GeV, the latter accounts for $\sim 70\% ~(90\%)$ of the total. In the present study, we have neglected the contribution from $\Xi_c$ states, since their fraction of total charm cross section is not significant~\cite{He:2019tik}.

The nuclear modification factor of a charm-hadron is defined as
\begin{equation}
\label{RAA}
R_{\rm AA}(p_T)=\frac{dN_{\rm AA}/dp_Tdy}{N_{\rm coll}/\sigma_{\rm NN}^{\rm in}d\sigma_{\rm pp}/dp_Tdy},
\end{equation}
where the numerator is the absolute $p_T$ differential yield of the charm-hadron under consideration in heavy-ion collisions, $d\sigma_{\rm pp}/dp_Tdy$ is the charm-hadron $p_T$ differential cross section in $pp$ collisions~\cite{He:2019tik} at the same colliding energy, and $N_{\rm coll}$ and $\sigma_{\rm NN}^{\rm in}$ are the binary collision number of the considered centrality bin and the inelastic nucleon-nucleon cross section, respectively. The nuclear modification factors of each charm-hadron species (compared to ALICE measurements) and of their corresponding charm muons are shown in Fig.~\ref{Charm-hadrons-muons_RAA}. An overall feature is a shift of the ``flow bump" in the well reproduced charm-hadron $R_{\rm AA}$'s toward lower $p_T$ as a consequence of the decay, which is more significant for the muons from $D_s^+$ and $\Lambda_c^+$ decays.

\begin{figure} [!t]
\includegraphics[width=0.99\columnwidth]{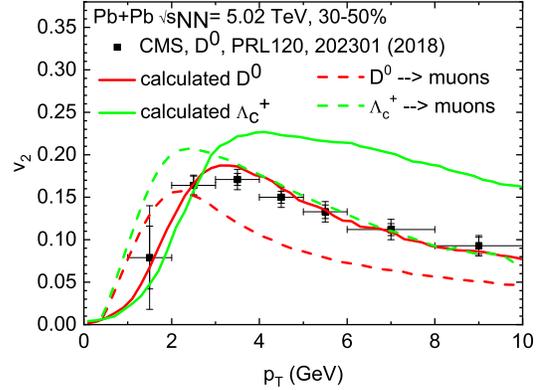}
\vspace{-0.15cm}
\caption{(Color online) $D^0$ and $\Lambda_c^+$ $v_2$ in $\sqrt{s}=5.02$\,TeV Pb-Pb collisions in the 30-50\% centrality bin at mid-rapidity, in comparison with those
of their decayed muons. The CMS data of $D^0$ $v_2$ are taken from~\cite{CMS:2017vhp}.}
\label{Charm-hadrons-muons_v2}
\end{figure}

The elliptic flow coefficient, defined as
\begin{equation}
\label{v2}
v_2(p_T)=\left\langle \frac{p_x^2 - p_y^2}{p_x^2 + p_y^2} \right\rangle ,
\end{equation}
characterizes the momentum anisotropy as a result of charm coupling with the medium through diffusion and hadronization. The $v_2$ of $D^0$ (compared to CMS measurements) and $\Lambda_c^+$ are shown in Fig.~\ref{Charm-hadrons-muons_v2} ($D^+$ and $D_s^+$ $v_2$ are similar to that of $D^0$), together with the $v_2$ of their decayed muons. An important observation here is that the $\Lambda_c^+$ $v_2$ is significantly larger than that of $D^0$ toward $p_T>3$\,GeV, as a result of three-body RRM incorporating SMCs that push the reach of recombination to higher $p_T$~\cite{He:2019vgs}. An immediate consequence is that the muons decayed from $\Lambda_c^+$ also have a significantly greater $v_2$ than the $D^0$ muons for $p_T>2$\,GeV, which may finally help increase the $v_2$ of total charm muons.

\section{Total charm leptonic observables vs data}
\label{sec_total_charm_leptons_vs_data}

\begin{figure} [!t]
\includegraphics[width=0.99\columnwidth]{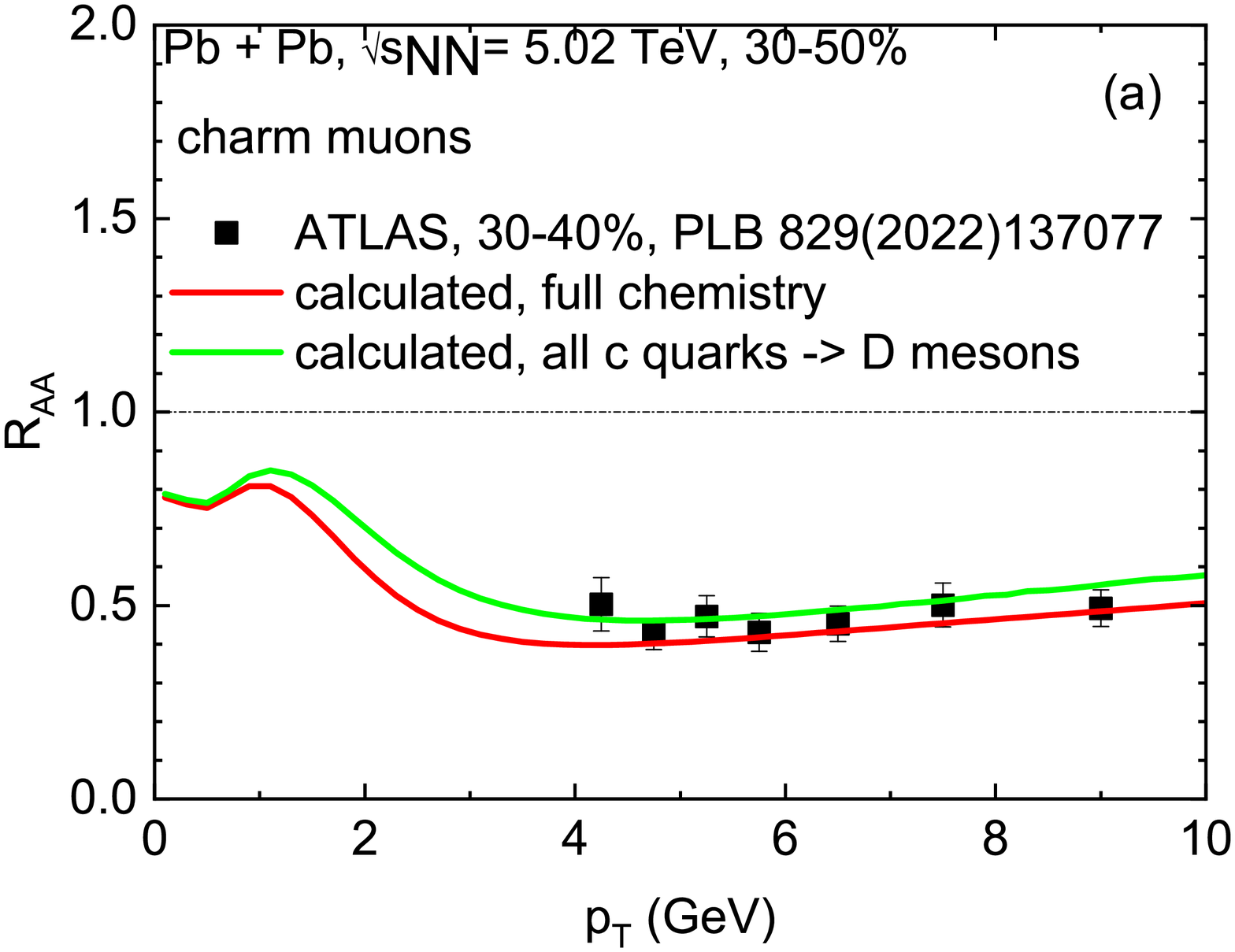}
\vspace{-0.15cm}
\includegraphics[width=0.99\columnwidth]{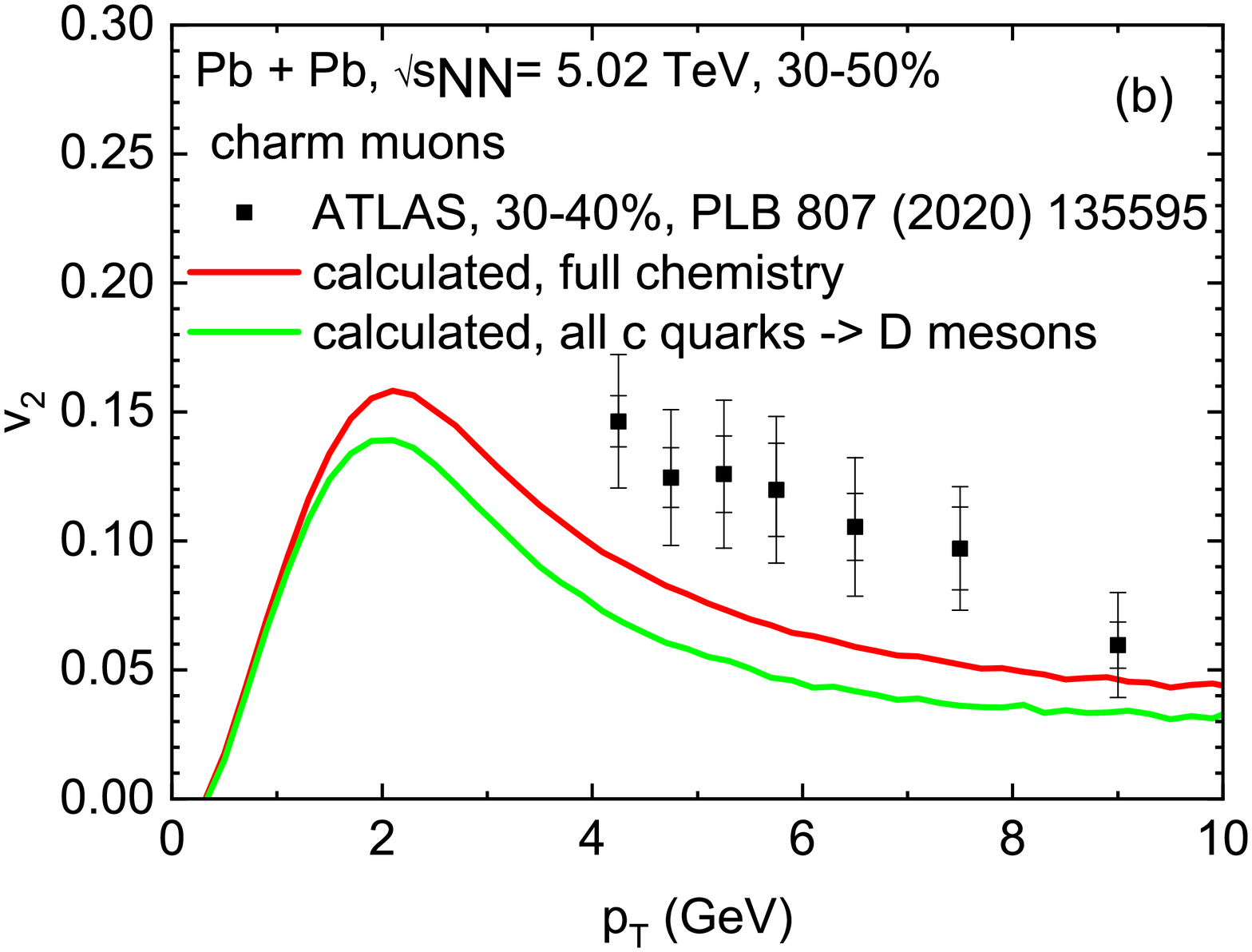}
\vspace{-0.15cm}
\caption{(Color online) $R_{\rm AA}$ and $v_2$ of total charm muons in $\sqrt{s}=5.02$\,TeV Pb-Pb collisions in the 30-50\% centrality bin at mid-rapidity from calculations taking account of full charm hadro-chemistry or from the scenario of charm quarks hadronizing into only $D$ mesons. ATLAS data are taken from~\cite{ATLAS:2020yxw,ATLAS:2021xtw} for charm muons with pseudorapidity cut $|\eta|<2.0$.}
\label{Charm-muons_RAA-v2_PbPb5TeV}
\end{figure}

Having analysed the observables of charm muons decayed from each charm-hadron species, we are now in a position to combine them together and show the results for the total charm muons. On the other hand, in previous studies of charm leptons within transport approaches~\cite{vanHees:2005wb,Gossiaux:2008jv,He:2011qa,He:2014cla,Song:2016rzw,Katz:2019fkc}, only $D$ mesons (denoting the sum of $D^0$ and $D^+$) were accounted for in the charm quark hadronization and therefore in the ensuing charm decayed leptons, ignoring in particular the role of the $\Lambda_c^+$ baryons (we have checked that $D_s^+/D^0$ enhancement as shown in Fig.~\ref{Charm-hadron_ratios} does not play a significant role in the observables of the final total charm muons, including their $v_2$, because $D_s^+$ decayed muons accounts for a rather minor fraction of the total as seen from Fig.~\ref{Charm-muons_spectra} and the $D_s^+$ $v_2$ is similar to that of the $D$ mesons). Here for the purpose of making out the pertinent charm hadro-chemistry effects through comparison, we have also conducted a calculation assuming all charm quarks are hadronized into $D$ mesons only, which are then subject to semileptonic decays to get the total charm leptons with an average branching ratio of $\sim 9.4$\% (obtained from an average between the branching ratios of $D^0$ and $D^+$ using the integrated ratio $D^+/D^0$ as pertinent weighs). More specifically, we use the charm quark recombination probability into $D$ mesons, $P_{D}^{\rm coal}(p_c^*)$, selfconsistently determined from the RRM formalism~\cite{He:2019vgs}, and renormalize it to unity at charm quark momentum $p_c^*=0$ (to ensure low momentum charm quarks are hadronized via recombination~\cite{He:2019vgs}), with the remaining $1-P_{D}^{\rm coal}(p_c^*)$ identified as the charm quark fragmentation probability into $D$ mesons. This follows the procedure done for the calculation with full charm hadro-chemistry, but in the latter, recombination probabilities of charm quarks into all charm-hadrons including the augmented charm-baryon excited states are added up and then renormalized~\cite{He:2019vgs}.

\begin{figure} [!t]
\includegraphics[width=0.99\columnwidth]{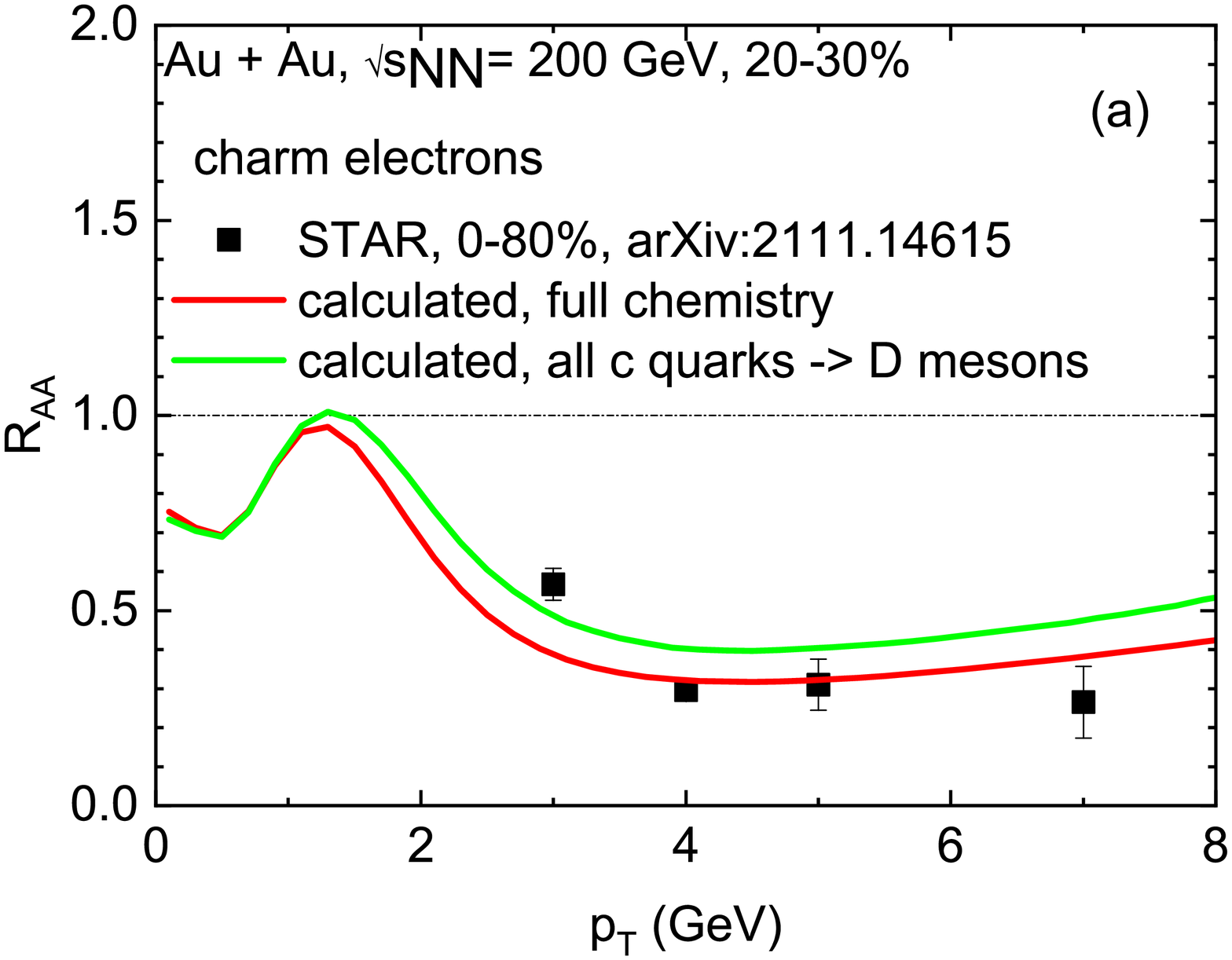}
\vspace{-0.15cm}
\includegraphics[width=0.99\columnwidth]{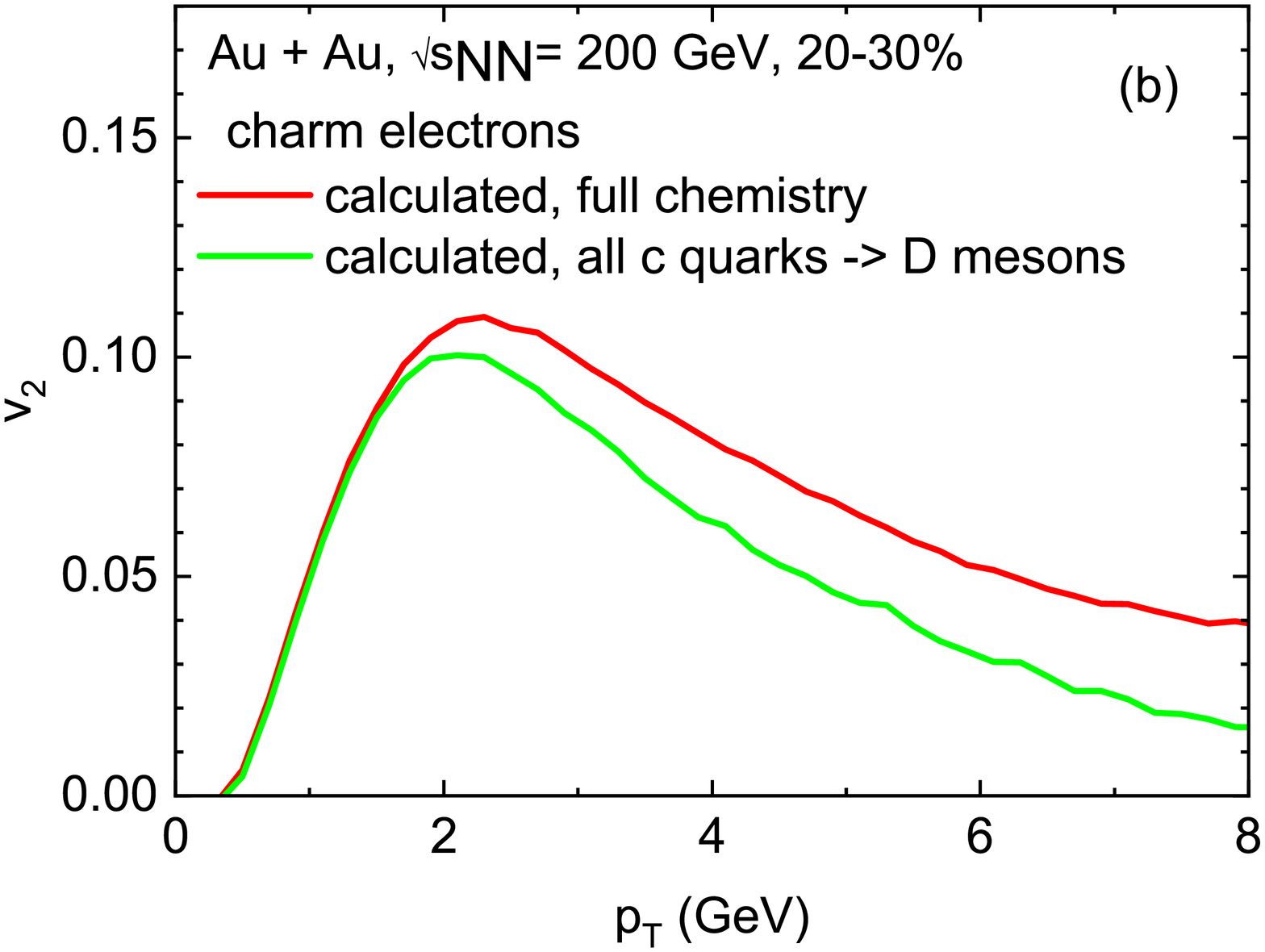}
\vspace{-0.15cm}
\caption{(Color online) $R_{\rm AA}$ and $v_2$ of total charm electrons in $\sqrt{s}=200$\,GeV Au-Au collisions in the 20-30\% centrality bin at mid-rapidity from calculations taking account of full charm hadro-chemistry or from the scenario of charm quarks hadronizing into only $D$ mesons. STAR data are taken from~\cite{STAR:2021uzu}.}
\label{Charm-electros_RAA-v2_AuAu200GeV}
\end{figure}

The calculated $R_{\rm AA}$ and $v_2$ of the total charm muons are compiled in Fig.~\ref{Charm-muons_RAA-v2_PbPb5TeV} in comparison with the ATLAS measurements for $\sqrt{s_{\rm NN}}=5.02$\,TeV Pb-Pb collisions in the 30-50\% centrality bin at mid-rapidity. For consistency, for calculating the $R_{\rm AA}$ in the scenario of all charm quarks hadronizing into $D$ mesons only, we have adopted the same hadro-chemistry in $pp$ collisions as in Pb-Pb collisions; {\it i.e.}, in $pp$ collisions all charm quarks are fragmented into only $D$ mesons whose $p_T$ spectrum as baseline is normalized to the realistic total charm cross section. One can read from the upper panel of Fig.~\ref{Charm-muons_RAA-v2_PbPb5TeV}, that taking full account of the charm hadro-chemistry only causes a mild change in the nuclear modification factor of the total charm muons, relative to the scenario of hadronizing charm quarks into $D$ mesons only. Both results are within the error bars of the ATLAS data that is currently limited to $p_T>4$\,GeV.

A more pronounced change (increase) in the $v_2$ of the total charm muons is seen upon full consideration of charm hadro-chemistry, as demonstrated in the lower panel of Fig.~\ref{Charm-muons_RAA-v2_PbPb5TeV}. This is mostly caused by the significantly greater $v_2$ of $\Lambda_c^+$ and its decayed muons than the $D$-meson counterparts, cf. Fig.~\ref{Charm-hadrons-muons_v2}. However, this increase is not as large as that seen in Fig.~\ref{Charm-hadrons-muons_v2}, since the total charm muons are dominated by those decayed from $D$ mesons and the contribution from $\Lambda_c^+$ accounts only for a minor fraction at intermediate $p_T$, as demonstrated in Fig.~\ref{Charm-muons_spectra}. The pronounced increase in the $v_2$ of the total charm muons as a result of full account of the charm hadro-chemistry seems to be supported by the large value of the ATLAS measurement in the 30-40\% centrality, although the present result is still significantly below the ATLAS data in nearly whole measured $p_T$ range.

The same calculations have been performed for $\sqrt{s_{\rm NN}}=200$\,GeV Au-Au collisions in the 20-30\% centrality bin as a proxy for the minimum bias collisions~\cite{He:2019vgs}. The results for the total charm electrons are shown in Fig.~\ref{Charm-electros_RAA-v2_AuAu200GeV}. The calculated $R_{\rm AA}$ from full consideration of charm hadro-chemistry is comparable to the STAR data in 0-80\% centrality, and the corresponding $v_2$ is also significantly greater than the result from the scenario of hadronizing charm quarks into $D$ mesons only.

\section{Summary}
\label{sec_summary}
In this work, inspired by the recent measurements of charm leptons by ATLAS and STAR experiments in Pb-Pb and Au-Au collisions, we have investigated the charm hadro-chemistry effects on the charm leptonic observables within our recently developed transport approach for charm probes~\cite{He:2019vgs}. Our study demonstrates that, because the total charm leptons spectrum is still dominated by the ones decayed from $D$ mesons even after taking into account full charm hadro-chemistry, the total charm leptons' nuclear modification factor does not change much as compared to the scenario of hadronizing all charm quarks into $D$ mesons only, which has been widely adopted by transport approaches when computing heavy flavor leptons. Yet the total charm leptons' elliptic flow acquires a pronounced increase because of the inclusion of the $\Lambda_c^+$ baryons that have significantly greater collectivity as a result of the three-body recombination, rendering the computed total charm muons $v_2$ closer to ATLAS data.

Going beyond the charm sector, bottom electrons $v_2$ has also recently been measured for the first time by ALICE experiment in Pb-Pb collisions at $\sqrt{s_{\rm NN}}=5.02$\,TeV~\cite{ALICE:2020hdw}. Calculated results from transport models~\cite{Nahrgang:2013xaa,Song:2016rzw,Ke:2018tsh} that assumed substantial interactions of bottom quark with the QGP and hadronized bottom quarks into $B$ mesons only significantly underestimated the bottom electron $v_2$ at intermediate $p_T$~\cite{ALICE:2020hdw}, signalling the potentially significant role of $\Lambda_b$ baryons (through its semileptonic decays) that are expected to have larger collectivity than $B$ mesons. This calls for full and controlled computation of bottom hadro-chemistry in transport approach like for the charm sector~\cite{He:2019vgs}, which will be addressed in our near-future work.

\acknowledgments This work was supported by NSFC grant 12075122.

\end{document}